\documentclass[aps,apl,twocolumn,groupedaddress,amsmath]{revtex4}
\usepackage[latin1]{inputenc}    
\usepackage{graphicx}   

\begin{document}
\title{Time-resolved detection of single-electron interference}
\author{S.~Gustavsson}
\email{simongus@phys.ethz.ch}
 \author{R.~Leturcq}
 \author{M.~Studer}
 \author{T.~Ihn}
 \author{K.~Ensslin}
 \affiliation {Solid State Physics Laboratory, ETH Z\"urich, CH-8093 Z\"urich,
 Switzerland}
\author{D. C.~Driscoll}
\author{A. C.~Gossard}
\affiliation{Materials Departement, University of California, Santa
Barbara, CA-93106, USA}

\date{\today}

\begin{abstract}
We demonstrate real-time detection of self-interfering electrons in
a double quantum dot embedded in an Aharonov-Bohm interferometer,
with visibility approaching unity.
We use a quantum point contact as a charge detector to perform
time-resolved measurements of single-electron tunneling.
With increased bias voltage, the quantum point contact exerts a
back-action on the interferometer leading to decoherence.
We attribute this to emission of radiation from the quantum point
contact, which drives non-coherent electronic transitions in the
quantum dots.
\end{abstract}

\maketitle

\begin{figure}[t] \centering
 \includegraphics[width=\columnwidth]{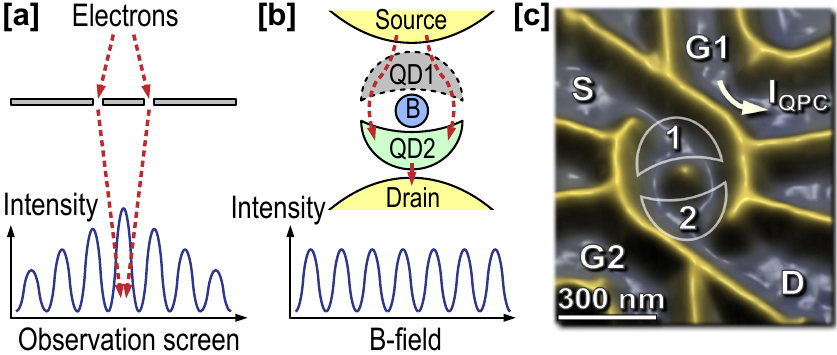}
 \caption{(a) Setup of a traditional double-slit experiment. Electrons passing
 through the two slits give rise to interference pattern on the observation screen.
 (b) Schematic drawing of the setup used for measuring single-electron Aharonov-Bohm interference.
 Electrons are injected from the source lead, tunnel through QD1 and end up in
 QD2, where they are detected.
 The interference pattern is due to the applied
 B-field, which introduce a phase difference between the left
 and right arm connecting the two quantum dots.
 (c) Double quantum dot used in the experiment. Yellow lines are
 written with a scanning force microscope on top of a semiconductor
 heterostructure and represent the potential landscape for the
 electrons. The
 QDs (marked by 1 and 2) are connected by two separate arms,
 allowing partial waves taking different paths to interfere.
 The current in the nearby QPC ($I_\mathrm{QPC}$) is used to monitor the
 electron population in the system.
 }
\label{fig:fig1}
\end{figure}

One of the cornerstone concepts of quantum mechanics is the
superposition principle as demonstrated in the double-slit
experiment \cite{young:1804}.
The partial waves of individual particles passing a double slit
interfere with each other. The ensemble average of many particles
detected on a screen agrees with the interference pattern calculated
using propagating waves [Fig.~\ref{fig:fig1}(a)]. This has been
demonstrated for photons, electrons in vacuum \cite{jonsson:1961,
tonomura:1989} as well as for more massive objects like
$C_{60}$-molecules \cite{arndt:1999}.
The Aharonov-Bohm (AB) geometry provides an analogous experiment in
solid-state systems \cite{aharonov:1959}. Partial waves passing the
arms of a ring acquire a phase difference due to a magnetic flux,
enclosed by the two paths [Fig.~\ref{fig:fig1}(b)]. Here we
demonstrate the self-interference of individual electrons in a
sub-micron Aharonov-Bohm interferometer. The interference pattern is
obtained by counting individual electrons passing through the
structure.

We first discuss the experimental conditions necessary for observing
single-electron AB interference. We make use of a geometry
containing two quantum dots (QD) within the AB-ring.
Figure~\ref{fig:fig1}(c) shows the structure with two QDs (marked by
1 and 2) tunnel-coupled via two separate barriers. The sample was
fabricated by local oxidation \cite{fuhrer:2004} of the surface of a
GaAs/AlGaAs heterostructure containing a two-dimensional electron
gas 34 nm below the surface. More details about the structure are
given in Ref.~\cite{gustavssonPRL:2007}. Following the sketch in
Fig.~\ref{fig:fig1}(b), electrons are provided from the source lead,
tunnel into QD1 and pass on to QD2 through either of the two arms.
Upon arriving in QD2, the electrons are detected in real-time by
operating a near-by quantum point contact (QPC) as a charge detector
\cite{field:1993}. Coulomb blockade prohibits more than one excess
electron to populate the structure, implying that the first electron
must leave to the drain before a new one can enter. This enables
time-resolved operation of the charge detector and ensures that we
measure interference due to individual electrons.

\begin{figure}[t]
\centering
 \includegraphics[width=\columnwidth]{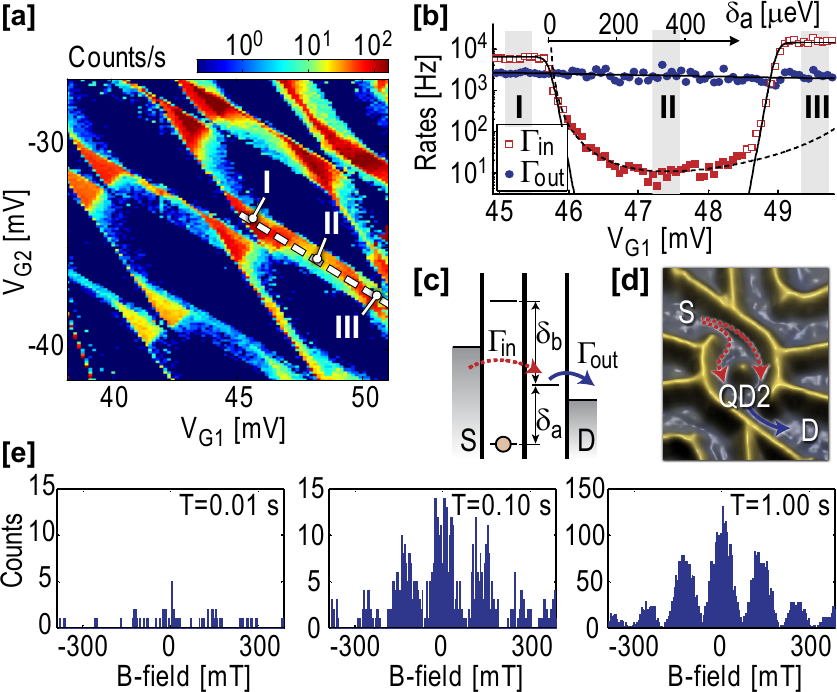}
 \caption{(a) Charge stability diagram of the double quantum dot,
 recorded by counting electrons entering and leaving the structure.
 The data was taken at bias voltage $V_\mathrm{b}=600~\mu\mathrm{V}$ and  $B=0~\mathrm{T}$.
 (b) Tunneling rates for electrons entering (red) an leaving (blue) the DQD,
 measured along the dashed white line in (a). The upper x-axis shows $\delta_a$, the potential
 difference between the state in QD2 and the occupied state of QD1. The solid
 lines are tunneling rates expected from sequential tunneling,
 while the dashed line is a fit to the cotunneling model of Eq.~(\ref{eqCot}).
 Parameters are given in the text. The data was taken with $B=340~\mathrm{mT}$.
 (c) Energy level configuration of the DQD at the point marked by II in (a,~b). Electron transport from source to
 QD2 is possible by means of cotunneling.
 (d) Schematic drawing of the cotunneling process.
 (e) Number of electrons arriving at QD2 within the fixed period of
 time indicated in the upper-right corner, measured as a function of magnetic field. The data was taken at point II in
 (a). The count rate shows an oscillatory pattern with a visibility higher than $90\%$.
 }
\label{fig:fig2}
\end{figure}

To avoid dephasing, the electrons should spend a time as short as
possible on their way from source to QD2. This is achieved by
raising the electrochemical potential of QD1 so that electrons in
the source lead lack an energy $\delta$ required for entering QD1
[see Fig.~\ref{fig:fig2}(c)]. The time-energy uncertainty principle
still allows electrons to tunnel from source to QD2 by means of a
second order process. The electron dwell time in QD1 is then limited
to a short time scale set by the uncertainty relation, with $t =
\hbar/\delta$ \cite{singleCharge:1992}.

The charge detector is implemented by tuning the QPC conductance
close to $0.5 \times 2e^2/h$. At this point the QPC conductance is
highly sensitive to changes in its electrostatic surroundings,
allowing it to be used to detect single electrons tunneling into or
out of the QD in real time \cite{vandersypen:2004, schleser:2004,
fujisawa:2004}. The QPC conductance was measured by applying a
d.c.~bias voltage over the QPC, $V_\mathrm{QPC} =250~\mathrm{\mu
V}$, and monitoring the current $I_\mathrm{QPC}$. The charge
detection technique allows the tunneling rates for electrons
entering and leaving the double QD (DQD) to be determined separately
\cite{gustavsson:2005, naaman:2006}.

In the experiment, we apply appropriate gate voltages to tune the
tunneling rates between the DQD and the source and drain leads to
values below $15~\mathrm{kHz}$.
The tunnel couplings between the QDs are set to a few GHz. The
interdot transitions are too fast to be detected with the bandwidth
of the charge detector ($\Gamma_\mathrm{det.} = 20~\mathrm{kHz}$),
but the coupling energy can still be determined from charge
localization measurements \cite{dicarlo:2004}.

Figure~\ref{fig:fig2}(a) shows the charge stability diagram of the
DQD, measured by counting electrons entering and leaving the
structure within a fixed period of time. The data was taken with
$600~\mathrm{\mu V}$ bias applied between source and drain. The
hexagon pattern together with the triangles of electron transport
appearing due to the applied bias are well-known characteristics of
DQD systems \cite{vanderwiel:2002}.
Between the triangles, there are band-shaped regions with weak but
non-zero count rates where electron tunneling is expected to be
suppressed due to Coulomb blockade. The finite count rate in these
regions can be attributed to electron tunneling involving virtual
processes \cite{gustavssonCot:2008}.

To investigate these processes in more detail, we follow the lines
of Ref.~\cite{gustavssonCot:2008} and plot the rates for electrons
tunneling into and out of the DQD measured along the dashed line in
Fig.~\ref{fig:fig2}(a). The result is shown in
Fig.~\ref{fig:fig2}(b). Going along the dashed line in
Fig.~\ref{fig:fig2}(a) corresponds to lowering the electrochemical
potential of QD1 while keeping the potential of QD2 constant. In the
region marked by I, electrons tunnel sequentially from the source
into QD1, continue from QD1 to QD2 and finally leave QD2 to the
drain lead. Proceeding to point II in Fig.~\ref{fig:fig2}(a,~b), the
electrochemical potential of QD1 is lowered and an electron is
trapped in QD1 [see sketch in Fig.~\ref{fig:fig2}(c)]. The electron
lacks an energy $\delta_a$ to leave to QD2, but because of
time-energy uncertainty there is a time-window of length
$\sim\!\hbar/\delta_a$ within which tunneling from QD1 to QD2
followed by tunneling from the source into QD1 is possible without
violating energy conservation. An analogous process is possible
involving the next unoccupied state of QD1, occurring on timescales
$\sim\! \hbar/\delta_b$. The processes correspond to electron
\emph{cotunneling} from the source lead to QD2. By continuing to
point III, the unoccupied state of QD1 is shifted into the bias
window and electron transport is again sequential.

The solid lines in Fig.~\ref{fig:fig2}(b) show the tunneling rates
expected from sequential tunneling \cite{kouwenhoven:1997}. The fit
gives the tunnel couplings between source and the occupied
($\Gamma_\mathrm{Sa}$)/unoccupied ($\Gamma_\mathrm{Sb}$) states of
QD1, with $\Gamma_\mathrm{Sa} = 6.4~\mathrm{kHz}$ and
$\Gamma_\mathrm{Sb} = 14~\mathrm{kHz}$.
In the cotunneling regions we fit the data to an expression
involving the sum of the two cotunneling processes
\cite{singleCharge:1992, gustavssonCot:2008}:
\begin{equation}\label{eqCot}
  \Gamma_\mathrm{cot} = \Gamma_\mathrm{Sa} t_a^2/\delta_a^2
  + \Gamma_\mathrm{Sb} t_b^2/\delta_b^2.
\end{equation}
Here, $t_a$, $t_b$ are the tunnel couplings between the
occupied/unoccupied states in QD1 and the state in QD2. The values
for $\Gamma_\mathrm{Sa}$ and $\Gamma_\mathrm{Sb}$ are taken from
measurement in the sequential regimes. The dashed line in
Fig.~\ref{fig:fig2}(b) shows the results of Eq.~(\ref{eqCot}), with
fitting parameters $t_a=8.3~\mathrm{\mu eV}$ and $t_b=13~\mathrm{\mu
eV}$. These values are in good agreement with values obtained from
the charge localization measurements. We emphasize that
Eq.~(\ref{eqCot}) is valid only if $\delta_a, \delta_b \gg t_a, t_b$
and if sequential transport is sufficiently suppressed, i.e. in the
range $46~\mathrm{mV} < V_\mathrm{G1} < 48.6~\mathrm{mV}$ of
Fig.~\ref{fig:fig2}(b).

Coming back to the sketch of Fig.~\ref{fig:fig1}(b), we note that
the cotunneling configuration of case II in Fig.~\ref{fig:fig2}(a-c)
is ideal for investigating the Aharonov-Bohm effect for single
electrons. Due to the low probability of the cotunneling process,
the source lead provides low-frequency injection of single electrons
into the DQD. The injected electrons cotunnel through QD1 into QD2
on a timescale $t \sim \hbar/\delta \sim 1~\mathrm{ps}$ much shorter
than typical decoherence times of the system \cite{eisenberg:2002}.
This ensures that phase coherence is preserved. Finally, the
electron stays in QD2 for a time long enough to be registered by the
finite-bandwidth charge detector. The tunneling processes are
visualized in Fig.~\ref{fig:fig2}(d).

To proceed, we tune the system to case II of
Fig.~\ref{fig:fig2}(a,~b) and count electrons as a function of
magnetic field. Figure~\ref{fig:fig2}(e) shows snapshots of the
number of electrons arriving in QD2 after three different times. The
electrons travel one-by-one through the system but still build up a
well-pronounced interference pattern with period $130~\mathrm{mT}$.
This corresponds well to one flux quantum $\Phi = h/e$ penetrating
the area enclosed by the two paths.
The visibility of the AB-oscillations is higher than $90\%$, which
is a remarkably large number demonstrating the high degree of phase
coherence in the system. We attribute the high visibility to the
short time available for the cotunneling process \cite{sigrist:2006}
and to strong suppression of electrons being backscattered in the
reverse direction, which is otherwise present in AB-experiments.
Another requirement for the high visibility is that the two tunnel
barriers connecting the QDs are carefully symmetrized. The overall
decay of the maxima of the AB-oscillation with increasing B is
probably due to magnetic field effects on the orbital wavefunctions
in QD1 and QD2.

\begin{figure}[t]
\centering
 \includegraphics[width=\columnwidth]{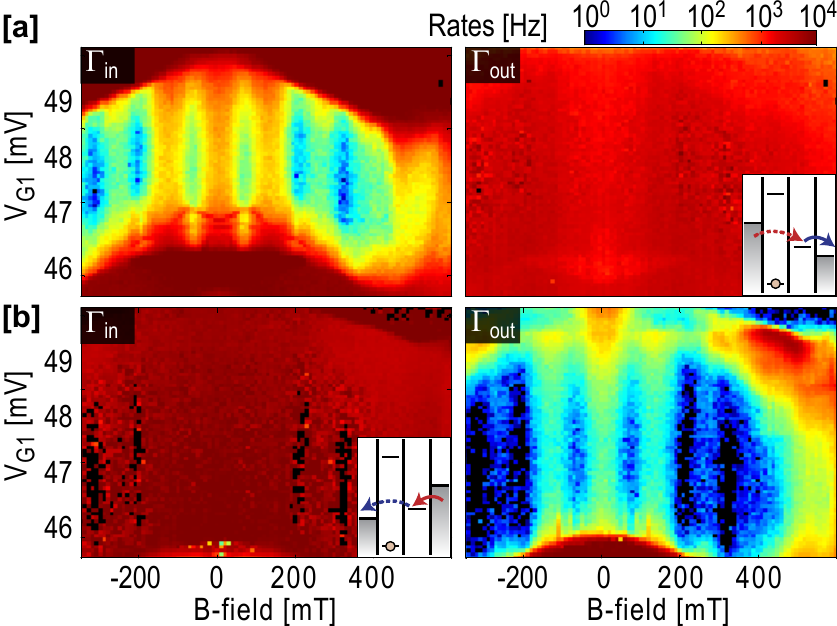}
 \caption{(a) Tunneling rates for electrons entering
 ($\Gamma_\mathrm{in}$) and leaving ($\Gamma_\mathrm{out}$) the DQD,
 measured versus electrochemical potential of QD1 and magnetic field.
 The y-axis corresponds to sweeps along the dashed line in
 Fig.~2(a). Within the cotunneling region,
 $\Gamma_\mathrm{in}$ shows clear B-field periodicity, while
 $\Gamma_\mathrm{out}$ remains constant. This is in agreement with
 the picture where only the electrons tunneling from source to QD2
 encircle the Aharonov-Bohm ring, while electrons leaving to drain
 remains unaffected by the applied B-field. (b) Same as (a), but
 with reverse bias over the DQD. Here, the roles of
$\Gamma_\mathrm{in}$ and $\Gamma_\mathrm{out}$ are inverted.
 }
\label{fig:fig3}
\end{figure}

Figure~\ref{fig:fig3}(a) shows the separate rates for electrons
tunneling into and out of the DQD as a function of magnetic field.
The y-axis corresponds to the dashed line in Fig.~\ref{fig:fig2}(a),
i.e., to the energy of the states in QD1. The measurement shows a
general shift of the DQD energy with the applied B-field, which we
attribute to changes of the orbital wavefunctions in the individual
QDs. Within the cotunneling region, $\Gamma_\mathrm{in}$ shows
well-defined B-periodic oscillations. At the same time,
$\Gamma_\mathrm{out}$ is essentially independent of the applied
field. This is expected since $\Gamma_\mathrm{out}$ measures the
rate at which electrons leave QD2 to the drain, which occurs
independently of the magnetic flux passing through the AB-ring [see
Fig.~\ref{fig:fig2}(c,~d)]. In Fig.~\ref{fig:fig3}(b), the bias over
the DQD is reversed. This inverts the roles of $\Gamma_\mathrm{in}$
and $\Gamma_\mathrm{out}$ so that $\Gamma_\mathrm{out}$ corresponds
to the cotunneling process. Consequently $\Gamma_\mathrm{out}$ shows
B-periodic oscillations while $\Gamma_\mathrm{in}$ remains
unaffected. In the black regions seen in Fig.~\ref{fig:fig3}(b) no
counts were registered within the measurement time of three seconds
due to strong destructive interference for the tunneling-out
process. As a consequence we could not determine the tunneling rates
in those regions.

\begin{figure}[t] \centering
 \includegraphics[width=\columnwidth]{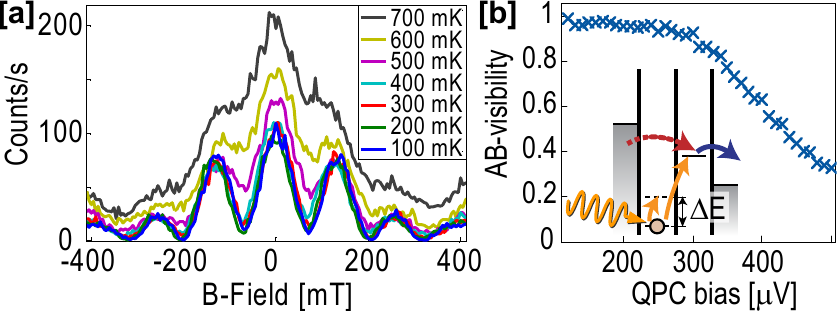}
 \caption{(a) Aharonov-Bohm (AB) oscillations measured at different
 temperatures. At $\sim 400~\mathrm{mK}$, the visibility of the oscillations drops drastically.
 The data was taken along the dashed line in (b).
 (b) Amplitude of AB-oscillations measured at different QPC bias.
 The inset shows a few photon absorbtion processes that are possible at large QPC bias.
 }
\label{fig:fig4}
\end{figure}

In Fig.~\ref{fig:fig4}(a), we investigate how the AB-oscillations
are influenced by elevated temperatures.
The dephasing of open QD systems is thought to be due to
electron-electron interaction \cite{altshuler:1997}, giving
dephasing rates that depend strongly on temperature
\cite{huibers:1998}. Figure~\ref{fig:fig3}(a) shows the temperature
dependence of the AB oscillations in our system.
The amplitude of the oscillations remains almost unaffected up to
$\sim\!400~\mathrm{mK}$, indicating that the coherence is not
affected by temperature until the thermal energy becomes comparable
to the single-level spacing of the QDs. We conclude that the
decreased visibility at higher temperatures is due to an increase in
thermal fluctuations of the QD population.

Decoherence can also occur because of interactions with the
environment.
In the experiment, we use the current in the QPC to detect the
charge distribution in the DQD. In principle, the QPC could also
determine whether an electron passed through the left or the right
arm of the ring, thus acting as a which-path detector
\cite{buks:1998, nederPRL:2007}. If the QPC were to detect the
electron passing in one of the arms, the interference pattern should disappear.
In Fig.~\ref{fig:fig4}(b), we show the visibility of the
AB-oscillations as a function of bias on the QPC. The visibility
remains unaffected up to $V_\mathrm{QPC} \sim \! 250~\mathrm{\mu
eV}$, but drops for higher bias voltages.

We argue that the reduced visibility is not due to which-path
detection. At $V_\mathrm{QPC} = 400~\mathrm{\mu V}$, the current
through the QPC is approximately $10~\mathrm{nA}$. This gives an
average time delay between two electrons passing the QPC of
$e/I_\mathrm{QPC} \sim \! 16~\mathrm{ps}$. Since this is ten times
larger than the typical cotunneling time, it is unlikely that the
electrons in the QPC are capable of performing an effective
which-path measurement.
Instead, we attribute the decrease of the AB-visibility to processes
where the DQD absorbs photons emitted from the QPC. Previous work
has shown that such processes may indeed excite an electron from one
QD to the other, as long as the energy difference between the QDs is
lower than the energy provided by the QPC bias
\cite{gustavssonPRL:2007}.
The radiation of the QPC may also drive transitions within the
individual QDs, thus putting one of the QDs into an excited state
\cite{onac:2006b, gustavssonNWQPC:2008}. A few absorbtion processes
are sketched in the inset of Fig.~\ref{fig:fig4}(b).

As long as the QPC bias is lower than both the DQD detuning
($\delta=400~\mathrm{\mu eV}$) and the single-level spacing of the
individual QDs ($\Delta E \sim \!200~\mathrm{\mu eV}$), the AB
visibility in Fig.~\ref{fig:fig4}(b) is close to unity.
When raising the QPC bias above $\Delta E$, we start exciting the
individual QDs. With increased QPC bias, more states become
available and the absorption process becomes more efficient. This
introduces new virtual paths for the cotunneling process. Since the
different paths may interfere destructively, the interference
pattern is eventually washed out.
In this way, the QPC has a physical back-action on the measurement
which is different from informational back-action
\cite{sukhorukov:2007} and which-path detection previously
investigated \cite{buks:1998,nederPRL:2007}.

In summary, we have demonstrated interference of single electrons in
a solid state environment. Such experiments have so far been limited
to photons or massive particles in a high-vacuum environment in
order to decouple the quantum mechanical degrees of freedom as much
as possible from the environment. Our experiments demonstrate the
exquisite control of modern semiconductor nanostructures which
enables interference experiment at the level of single
quasi-particles in a solid state environment.
Once extended to include spin degrees of freedom \cite{loss:2000}
such experiments have the potential to facilitate entanglement
detection \cite{saraga:2003} or investigate the interference of
particles \cite{nederNature:2007} originating from different
sources.
\\


\end{document}